\documentclass[prl,aps,amssymb,twocolumn,superscriptaddress]{revtex4}
\usepackage{psfig}

\def\ket#1{\vert#1\rangle}
\def\bra#1{\langle#1\vert}
\def\braket#1#2{\langle#1\vert#2\rangle}
\def\br{{\bf r}}
\def\bk{{\bf k}}
\def\im{{\rm Im}}
\def\re{{\rm Re}}
\def\pmb#1{\setbox0=\hbox{#1}%
 \hbox{\kern-.025em\copy0\kern-\wd0
 \kern.05em\copy0\kern-\wd0
 \kern-0.025em\raise.0433em\box0} }

\def\taub{{\pmb{$\tau$}}}

\begin{document}

\title{Complex band structure with ultrasoft pseudopotentials: fcc Ni 
and Ni nanowire}

\author{A. Smogunov}
\affiliation{SISSA, Via Beirut 2/4, 34014 Trieste (Italy)}
\affiliation{INFM, Democritos Unit\`a di Trieste Via Beirut 2/4, 34014
Trieste (Italy)}
\author{A. Dal Corso}
\affiliation{SISSA, Via Beirut 2/4, 34014 Trieste (Italy)}
\affiliation{INFM, Democritos Unit\`a di Trieste Via Beirut 2/4, 34014
Trieste (Italy)}
\author{E. Tosatti}
\affiliation{SISSA, Via Beirut 2/4, 34014 Trieste (Italy)}
\affiliation{INFM, Democritos Unit\`a di Trieste Via Beirut 2/4, 34014
Trieste (Italy)}
\affiliation{ICTP, Strada Costiera 11, 34014 Trieste(Italy)}

\date{\today}

\begin{abstract} 
We generalize to magnetic transition metals the approach proposed by Choi 
and Ihm for calculating the complex band structure of periodic systems,
a key ingredient for future calculations of conductivity of an open quantum
system within the Landauer-Buttiker theory. The method is implemented 
with ultrasoft pseudopotentials and plane wave basis set in a 
DFT-LSDA ab-initio scheme. 
As a first example, we present the complex band structure
of bulk fcc Ni (which constitutes the tips of a Ni nanocontact) and 
monatomic Ni wire (the junction between two tips).
Based on our results, we anticipate some features of the spin-dependent
conductance in a Ni nanocontact. 
\end{abstract}
\maketitle

\section{Introduction}
Recent conductance data of nanocontacts and break junctions in
magnetic transition metals such as Ni have shown interesting and 
partly unexpected results. 
While early conductance histograms for Ni at room temperature in air appeared 
basically structureless \cite{costa}, Oshima et al. \cite{oshima},
who worked in vacuum, at variable temperature, and with the possibility
of a magnetic field, found a minimal conductance step preferentially near 2
and 4 (in units of $g_0=e^2/h$, the conductance quantum per spin)
at RT and zero field, near 4 at 770 K and zero field, and near
3 (occasionally near 1) at RT with a field. Ono et al. \cite{ono}, 
reported again 2 for Ni in zero field, and 1 for Ni in a field. 
Break junction data by Yanson \cite{yanson} show a minimal conductance step 
of about 3.2 in zero field.
Despite the poor consistency between these data, it is clear that 
the lowest conductance step for a Ni nanocontact is anomalously small 
in comparison with the large expected number of $s+d$ conducting
channels, of the order of at least 8 \cite{our} for a monatomic contact.

In a previous paper \cite{our} we investigated the possibility that the 
smallness of the last conductance step observed for a Ni nanocontact 
may be caused by the fact that two tips are magnetized
in different directions forming a magnetization reversal (or domain
wall) precisely in the contact region. By replacing the contact with
a monatomic infinite tipless Ni wire we found that the number of
conducting channels (energy bands crossing the Fermi level) can be
significantly reduced from 8 (nonmagnetic wire) or 7 (ferromagnetic wire) to
2 as a magnetization reversal is built into the
wire. The two remaining conducting channels correspond to light $s-$electrons,
while the heavy $d-$electrons are confined in the regions of positive and
negative magnetization and cannot propagate along the wire due to the
magnetization reversal. 

The real nanocontact consisting of two macroscopic tips and an
atomic contact region is an
open quantum system and has no artificial periodic conditions such as those
in Ref.\cite{our}.
The influence of the tips themselves (not considered in the previous
model)  will be in general important for the
nanocontact conductance. So, in order 
to make quantitative conductance predictions 
one will need to adopt a more detailed approach 
comprising two semi-infinite metals (either thick wires or semi-infinite
crystals) connected by a neck (a short monatomic wire, or just a single
atom). Several theoretical methods are available to study the transport
properties of atomic scale conductors. Among these are methods 
based on nonequilibrium Green's functions combined with a localized
basis set \cite{taylor1,taylor2}.
The Landauer-Buttiker formula provides another simple way to calculate
the conductance of quantum systems connected to two tips. 
The conductance is expressed in terms of the transmission coefficient
at the Fermi level $E_F$. In order to compute the transmission coefficient
one must solve a scattering problem for the open quantum system.
The scattering approach has been applied by Lang and co-workers \cite{lang},
and Tsukada and co-workers\cite{hirose1,hirose2} using the jellium model for
the electrodes. The layer KKR approach has been used to study the 
spin-dependent tunneling in magnetic tunnel junctions \cite{lkkr}. 
Lately Wortmann and co-workers\cite{wortmann} have formulated the transfer
matrix approach based on computation of single-electron Green function
in the linearized augmented plane wave (LAPW) basis. We are interested
in a formulation based on plane waves and ultrasoft pseudopotentials,
a calculational technology which we are applying to other problems
involving transition metals \cite{dalcorso1} with great potential
for the future.
  
Recently Choi and Ihm \cite{choi} presented a first-principles plane wave 
based solution to the scattering problem with real atomic contacts
incorporating the Kleinman-Bylander-type pseudopotentials. 
We wish to generalize this scheme to deal with ultrasoft 
pseudopotentials \cite{vanderbilt} which adequately
describe the nuclei and core electrons of a transition metal such as Ni.
The first step, on which we focus in this paper is the complex band structure
of periodic system, consisting of both propagating and exponentially decaying
evanescent states. These generalized Bloch states will be needed to construct
the scattering state deep inside the tips. The evanescent states play a
significant role in contact, surface or interface phenomena where there is a
breaking  of translational symmetry in one direction. 

We present here the complex band structure of bulk Ni (which gives
information on the nanocontact tips) and of a monatomic Ni wire (the junction
between the two tips). Although this is clearly only the first step
towards a full theoretical description of the nanocontact conductivity, we
shall see that some insight can be  obtained already by examining the complex
bands. The importance of  complex band structure for conductance
predictions has been recently emphasized for magnetic tunnel junctions
\cite{mavrop} and molecular systems \cite{sankey}.
        
\section{Method}
We consider an open quantum system divided into three regions: 
left bulk contact ($z<0$), scattering region ($0<z<L$) and right bulk contact 
($z>L$) and we use a supercell geometry in the $xy$ direction perpendicular 
to the wire. 
A wave propagating at energy $E$ with $\bk=(\bk_\perp,k)$   
and incident from the left contact on the scattering region will form 
a scattering state $\Psi$ which due to the supercell geometry has 
a Bloch form in the $xy$ direction: 
\begin{equation} 
\Psi_{{\bf k}_\perp}(\br_{\perp}+{\bf R}_\perp,z;\sigma)=e^{i{\bf k}_\perp 
{\bf R}_\perp} \Psi_{{\bf k}_\perp}(\br_\perp,z;\sigma).
\label{uno} 
\end{equation} 
Different ${\bf k}_\perp$ do not mix and can be considered separately. 
 If there is no coupling between different spin polarizations in the 
 Hamiltonian the same conclusion is valid also for the spin index $\sigma$ 
 and we can omit both ${{\bf k}_\perp}$ and $\sigma$ from the following 
 formulas. 

Deep within the contacts the system becomes periodic also along the $z$
direction and one can determine generalized Bloch states with complex 
$k$ vector which
obey to the periodic condition:
\begin{equation} 
 \psi_k({\bf r}_\perp,z+d)=e^{i 
 k 
 d}\psi_k({\bf r}_\perp,z), 
\label{bc}
\end{equation}
where $d$ is the length of the periodic unit cell deep within the contacts. 
In this region the scattering state
$\Psi$ can be written as a combination 
of propagating and evanescent generalized Bloch states: 
\begin{displaymath} 
\Psi= \left\{\begin{array}{ll} 
\psi_k+\sum\limits_{k'\in L}r_{kk'} 
\psi_{k'} 
, & {\rm ~~~in~the~left~tip} 
\\ 
\quad\sum\limits_{k'\in R}t_{kk'} 
\psi_{k'} 
, & {\rm ~~~in~the~right~tip} 
\end{array} \right. 
\end{displaymath} 
where summation over $k'\in L~(k'\in R)$ includes the generalized Bloch states 
in the left (right) contact at energy $E$ which propagate or decay to the left 
(right). In the $xy$ plane the wave functions $\psi_{k'}$ obey the same Bloch 
condition as in Eq.(\ref{uno}). 
Thus to compute the transmission and reflection coefficients 
$\{t_{kk'},r_{kk'}\}$ one will need to find the complex band structure and
generalized Bloch states $\psi_k$ in the metallic contacts at the chosen
energy $E$.  

We determine here the complex band structure within Density Functional Theory 
(DFT) describing the atoms with ultrasoft pseudopotentials \cite{vanderbilt}.
We consider the generalized Bloch states which satisfy the single-particle 
Kohn-Sham equation 
\cite{laasonen} 
(atomic units $e^2=2m=\hbar=1$ are used): 
\begin{equation} 
E\hat S\ket{\psi_k}=\left[-\nabla^2+V_{\rm eff}+ 
\hat V_{NL}\right]\ket{\psi_k}, 
\end{equation} 
where $V_{\rm eff}$ is the effective local potential 
(see Ref.~\cite{vanderbilt,laasonen}), 
$V_{NL}$ is the nonlocal part of the ultrasoft pseudopotential: 
\begin{equation} 
\hat V_{NL}=\sum_{Imn}D^I_{mn}\ket{\beta^I_m} 
\bra{\beta^I_n}, 
\end{equation} 
constructed using the set of projector functions $\beta^I_m$ associated 
with atom $I$ and $\hat S$ is the overlap operator: 
\begin{equation} 
\hat S=1+\sum_{Imn}q^I_{mn}\ket{\beta^I_m} 
\bra{\beta^I_n}.
\end{equation}
The coefficients $q^I_{mn}$ are the integrals of 
the augmentation functions defined in Ref.~\cite{vanderbilt,laasonen}. 
It is convenient to rewrite Eq.~(3) in the form: 
\begin{eqnarray} 
E\ket{\psi_k}&=&\left[-\nabla^2+V_{\rm eff}\right]\ket{\psi_k} \nonumber\\
&+&\sum_{Imn}(D^I_{mn}-Eq^I_{mn})\ket{\beta^I_m}\braket{\beta^I_n} 
{\psi_k}. 
\label{sequation} 
\end{eqnarray} 
From this equation it can be seen that the scattering problem 
in the ultrasoft pseudopotential case is not substantially different from 
the norm conserving one (where $E$ does not appear on the r.h.s.) 
since the energy $E$ is fixed and is given as an 
external parameter. 

As in Ref.~\cite{choi} we can write the solution of 
the integro-differential Eq.~(\ref{sequation}) as  
a linear combination:   
\begin{equation} 
\psi_k(\br)=\sum_n a_{n,k}\psi_n(\br)+\sum_{Im} c_{Im,k} 
\psi_{Im}(\br), 
\label{tre} 
\end{equation} 
where $\psi_n$ are linearly independent solutions of the homogeneous equation: 
\begin{equation} 
E\psi_n(\br)=\left[-\nabla^2+V_{\rm eff}(\br)\right]\psi_n(\br), 
\end{equation} 
and $\psi_{Im}$ is a particular solution of the inhomogeneous equation: 
\begin{eqnarray} 
E\psi_{Im}(\br)&=&\left[-\nabla^2 
+V_{\rm eff}(\br)\right]\psi_{Im}(\br) \nonumber\\
&+&\sum_{{\bf R}_\perp} 
e^{i{\bf k}_\perp {\bf R}_\perp} 
\beta^I_m(\br-\taub^I-{\bf R}_\perp). 
\end{eqnarray} 
Both $\psi_n$ and $\psi_{Im}$ are $(x,y)$ periodic as in
Eq.~(\ref{uno}).  Summation over $Im$ in Eq.~(7) involves all the projectors in
the unit cell  and the coefficients $c_{Im,k}$ are determined by: 
 \begin{equation} 
 c_{Im,k}=\sum_n [D^I_{mn}-Eq^I_{mn}] 
 \int [\beta^I_n(\br-\taub^I)]^*\psi_k(\br)d^3r. 
 \end{equation} 
Note that the set 
of wave functions $\psi_n$ is infinite. In practice the 
expansion $\psi_n(\br)=\sum_{{\bf G}_\perp}\psi_n({\bf G}_\perp,z) 
e^{i({\bf k}_\perp+{\bf G}_\perp)\cdot\br_\perp}$ over two-dimensional 
plane waves is used. If one considers only $N_{2D}$ plane waves with 
${\bf G}_\perp^2\leq E_{\rm cut}$ the number of $\psi_n$ becomes finite 
and equals to $2N_{2D}$. As in Ref.\cite{choi} to find the functions $\psi_n$ 
we discretize the unit cell along the $z$ axis by dividing it 
into slices and replacing the $V_{\rm eff}(\br)$ in each slice 
by a $z$-independent potential. The $\psi_n$ in the slices will be a 
linear combination of two exponentials with coefficients obtained 
by a matching procedure.

The allowed values of $k$ for a given energy $E$ can be determined by 
imposing periodicity along $z$ (Eq.~(\ref{bc})) to the generalized 
Bloch function $\psi_k(\br)=\sum_{{\bf G}_\perp}\psi_k({\bf G}_\perp,z) 
e^{i({\bf k}_\perp+{\bf G}_\perp)\cdot\br_\perp}$ and to its $z$ derivative: 
\begin{equation} 
\psi_k({\bf G}_\perp,d)=e^{i 
k 
d} 
\psi_k({\bf G}_\perp,0).
\end{equation} 
\begin{equation} 
\psi'_k({\bf G}_\perp,d)=e^{i 
k 
d} 
\psi'_k({\bf G}_\perp,0).
\end{equation} 
Inserting Eq.~(7) into Eqs.~(10)-(12) one can show
that the last three equations are equivalent to the
generalized eigenvalue problem: 
\begin{equation}
AX=e^{i 
k 
d}BX. 
\end{equation} 
where $A$ and $B$ are matrices.
We solve this problem to obtain in general a complex $k$ and coefficients 
$X=\Big\{a_{n,k},c_{Im,k}\Big\}$ for the generalized Bloch state 
$\psi_k$ at a given energy $E$ and $\bk_\perp$. 

Our complex band structure calculations proceed in two steps. First we compute 
with a standard electronic structure code 
({\tt PWSCF}) \cite{pw} the ground state electronic structure of the system 
(infinite bulk; and, separately, infinite perfect nanowire)
and we obtain 
the effective local potential $V_{\rm eff}(\br)$ and the screened
coefficients $D^I_{mn}$~\cite{notapw}. 
In a second step, we use the potential $V_{\rm eff}(\br)$ and the screened
coefficients $D^I_{mn}$ to compute the complex band structure applying 
the scheme described above~\cite{notacond}.   

\section{Results and discussion}

As a first example, we studied the complex band structure of ferromagnetic 
bulk fcc Ni (Fig.1). 
The orientation of the Ni tips in the actual break junction experiments 
is not under control, and we have chosen to plot the complex bands appropriate 
for a $(001)$ orientation of the electrodes. 
We considered a tetragonal unit cell with two nickel atoms and 
the $z$ axis parallel to the tetragonal axis. 
Our calculated optimal equilibrium lattice constant of ferromagnetic fcc Ni is 
$a=3.42$ \AA\ (expt. $a=3.52$ \AA) and this value has been 
used for the band 
structure calculation. 
At this volume the magnetic moment per atom is 0.59 $\mu_B$ (expt. 0.61
$\mu_B$) \cite{dalcorso2}. 
In the scattering problem we took ${\bf k}_\perp=0$ for the purpose
of examplification. 
Due to the symmetry of the Hamiltonian \cite{chang},
if at a fixed energy $E$ a real $k$ is a solution 
of the problem, also $-k$ is a solution (corresponding to right- and left- 
propagating waves). If a complex $k$ is a solution also $-k,k^*,-k^*$ 
are solutions \cite{chang} (states with $\im\ k>0$ ($\im\ k<0$) correspond to 
evanescent modes decaying along the positive (negative) $z$ direction). 
We can therefore show the complex band structure just in the 
region $\re\ k\geq0$ and $\im\ k\geq0$.   

\begin{figure}
\psfig{file=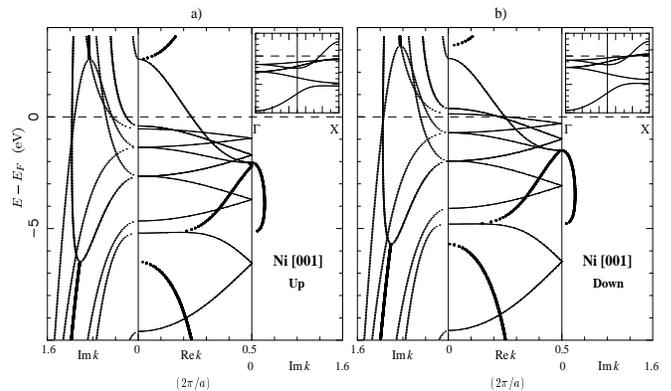,height=5.2cm,width=8.7cm,angle=0}
\caption[]
{
Complex band structure of fcc Ni along [001] direction at $\bk_\perp=0$
for the majority (spin up) and minority (spin down) states. Tetragonal
unit cell containing two atoms was used. Real
bands  (solid curves), complex bands with $\re\ k=0$ (dotted
curves) are plotted in the middle and left panels, respectively.
The real projection of complex bands (open circles) is plotted in the middle
panel, and the corresponding imaginary projection (open circles) is plotted
in the left or right panels. The band structures alog $\Gamma$-X direction
in the fcc Ni are shown on the insets.
}
\label{fig1}
\end{figure}

Ni is a ferromagnetic metal and we plot in Fig.~1a and 1b the bands 
corresponding to the majority spin (spin up) and minority spin (spin down) 
respectively. 
Figs.~1a and 1b are divided into three panels. The propagating bands with 
$\re\ k\ne0$ and $\im\ k=0$ are plotted with continuous lines in the central 
panel. 
Dotted lines in the left panel represent bands with purely imaginary 
$k$ ($\re\ k=0$ and $\im\ k\neq0$). The bands which 
correspond to complex $k$ with $\re\ k\neq0$ and $\im\ k\neq0$ are 
represented with open circles. They are shown with two branches which 
correspond to their projection in the real and imaginary planes. 
We chose arbitrarily to plot the projection on the imaginary plane 
in the left or in the right part of the figure in order to improve 
clarity. 
The band structure of fcc Nickel corresponding to real $k$ obtained 
with {\tt PWSCF} are shown in the 
inset of Fig.~1 along the ${\it\Gamma}-X$ direction. 
Since our cell has two atoms, the bands with real $k$ in Fig.1 should coincide 
with the one in the inset after folding about the middle vertical line. 
We indeed found excellent agreement between the real band structures in the two
calculations, with differences of the order of 0.005 eV  (invisible in the
figure). 

In the band structure of fcc Ni one can identify five 
narrow $3d$ bands and one broader $4s$ band. 
Along the $(001)$ direction the $s$ band mixes with the $d_{z^2}$ band. 
The spin up $d$ bands are fully occupied and only the $s-d_{z^2}$ band 
crosses the Fermi level. 
The down spin $d$ bands are upshifted in energy by a calculated exchange
splitting  of about 0.5 eV (exp. 0.33 eV). Therefore they are partially
unoccupied and  three $d$ states in addition to the $s-d_{z^2}$ state crosses
the Fermi level.  The complex bands shown in Fig.1 can be interpreted in terms
of general  theorems which describe their topological features \cite{chang}.
It  is known that whenever a band with real $k$ has a local extremum there is 
a complex band with opposite curvature and the same symmetry  which departs at
the same $\re\ k$ at right angles from the real plane. Complex $k$ bands
are parabolic or form loops joining real bands  of the same symmetry. 
One such loop in the complex plane joins $s-d_{z^2}$ bands connecting the 
maximum at $\re\ k=0.18~(2\pi/a)$ (very weakly
pronounced)   with the minimum at $k=0.5~(2\pi/a)$. 
Comparing the complex band structures for different spin polarizations one 
can notice that there are four propagating spin down bands at $E_F$ 
while for spin up polarization the $d$ bands cross the Fermi level only 
in the imaginary plane and are therefore evanescent (the smallest decay is 
$\im\ k=0.2~(2\pi/a)$). 

\begin{figure}
\psfig{file=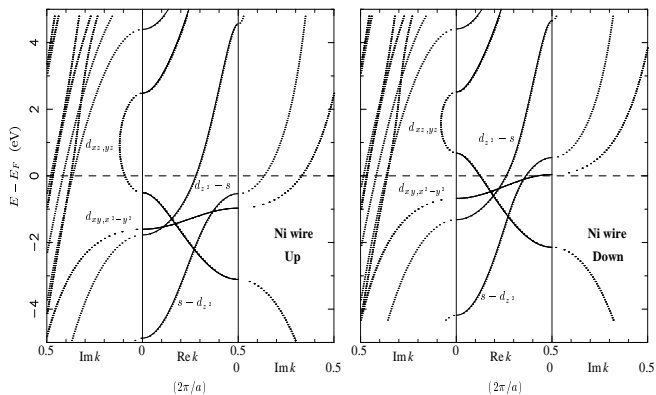,height=5.2cm,width=8.7cm,angle=0}
\caption[]
{
Complex band structure of monatomic Ni wire
for the majority (spin up) and minority (spin down) states.
Real bands  (solid curves), complex bands with $\re\ k=0$, and $\pi/a$ (dotted
curves) are plotted in the middle, left, and right panels, respectively.
Each band is labeled by its main atomic character.
}
\label{fig2}
\end{figure}

In Fig.~2 we show the complex band structure for a monatomic Ni 
wire ($a=2.12$ \AA). The supercell geometry and the other technical 
parameters used in this calculation were already discussed in
Ref.~\cite{our}.  The bands at real $k$ coincide (within 0.005 eV) with those
presented there and calculated with the {\tt PWSCF} code. 
Here all the occupated real bands have maxima or minima at the zone border 
and the complex bands are therefore either purely imaginary (they are shown
by dots in the left panel) or
have $\re\ k=\pi/a$ (they are shown by dots in the right panel). 
Bands at imaginary $k$ have a parabolic  dispersion except for the loop which
joins $d_{xz,yz}$ bands.  As in the case of bulk Ni at the Fermi energy 
there is only one real $s-d_{z^2}$ band of spin up polarization. All the 
$d$ states are decaying and the smallest decay ($d_{xz,yz}$ states) is
$\im k=0.08~(2\pi/a)$. On
the contrary 6 spin down bands with real $k$  cross the Fermi level and the
only decaying states crossing the Fermi energy  have a much larger imaginary
part $\im\ k=0.35~(2\pi/a)$. 

The asymmetry between energy bands at $E_F$ for different spin directions and
the existence of evanescent states with long decay lengths (small decay
parameters $\im k$) will play an important role in transport properties of the
magnetic nanocontacts. One can see, for example, that if there is no spin
reversal (two tips and the
wire are magnetized in the same direction) all propagating states coming from
the bulk Ni can get (though with some reflection at the junctions) through the
wire. On the other hand, if there is one spin reversal (situated either on 
one end of the wire or in the middle) the propagating $d$-states of one Ni tip
become decaying in another Ni tip so that they will be completely blocked.
The only open channels correspond to $s$-electrons (this
conclusion was arrived at \cite{our} considering a simple model of the Ni
nanocontact). And, finally, in the hypothetical case of two spin reversals
(when the wire has the opposite direction of magnetization with respect to
both tips) the $d$-electrons are only able to tunnel across the wire through
the evanescent states. The latter have an exponential decaying form $\psi_k\sim
e^{-z/\ell}$  with the decay length $\ell=7.96$ a.u. ($d_{xz,yz}$),
5.31 a.u. ($s-d_{z^2}$), and 2.0 a.u. ($d_{xy,x^2-y^2}$)
which should be compared with the distance between atoms in the Ni wire $a=4.0$ a.u. 
The shorter the wire, the larger the contribution of $d$-electrons to the total
conductivity. This result is similar to the spin-dependent tunneling
in magnetic tunnel junctions (see \cite{lkkr,mavrop}) although in our case 
it is not so crucial since two propagating $s$-channels always exist. 
All these ingredients will become crucial when one will 
solve the scattering problem in order to find the transmission coefficient.
It should be emphasized that the number of propagating channels $N$ give
the upper limit for conductivity, $G={e^2\over h} N$ with
unit transmission for all channels, while the scattering problem solution
may produce very small transmission coefficients for some propagating channels.
The channel number $N$ of a thick wire will diverge proportional to the wire
cross section, realizing the perfect conductance expected for a macroscopic 
case. 

\section{Conclusions}

We are working to generalize to magnetic transition metals the approach
proposed by Choi and Ihm for calculating the ballistic conductance of open
quantum systems within the Landauer-Buttiker theory.
In this paper we focused on the complex band structure of periodic systems,
which is the key ingredient for the wave-function matching between the
scattering region and the metallic electrodes. 
The method has been implemented for the first time with ultrasoft 
pseudopotentials and plane waves in a DFT-LSDA ab-initio scheme. 
We showed that it can be applied to both the bulk solid 
and to the idealized monatomic nanowire. As a first
application we calculated the complex band structure of fcc Ni and of a
ferromagnetic  Ni monoatomic wire and discussed qualitatively the
properties of a nickel nanocontact which can be inferred from the complex
band structure. Detailed calculations of conductance will be undertaken in
forthcoming work.

\section{Acknowledgments}

This work was partly sponsored by MIUR PRIN, by INFM (section F, G,
PRA NANORUB and ``Iniziativa Trasversale calcolo parallelo''), and
EU Contract ERBFMRXCT970155 (FULPROP). 
Some calculations were performed on the IBM-SP3 at CINECA,
Casalecchio (Bologna).


\begin{thebibliography}{25}
\bibitem {notapw} The calculations are based on local spin density 
approximation (LSDA)
with the Perdew-Zunger parametrization \cite{perdew} of exchange-correlation 
energy.  The nuclei and the core electrons are described by ultrasoft 
pseudopotentials with the parameters of Ref.~\cite{our}. 
The kinetic energy cut-offs of the plane wave basis set were 25 Ry 
and 300 Ry for the wave functions and the charge density, respectively. 
$\bk$-point sampling is done with 60 points in the irreduducible 
BZ for bulk Ni and with a uniform mesh 
of 40 $\bk$-points in the 1D BZ zone for the Ni wire. 
The integration up to the Fermi level was done by 
a standard broadening technique \cite{smearing} with the smearing 
parameter of 0.005 Ry.
\bibitem {notacond}
The 2D cut-off energy was taken as in electronic structure
calculations and the number of slices was 24 for bulk Ni and 15 for Ni wire
(in both cases the slice thickness is about 0.14 \AA) which proved 
sufficient to represent the shape of the potential
$V_{\rm eff}(\br)$.
\end{thebibliography}
\end{document}